\documentclass[prl,twocolumn,amsmath,amssymb]{revtex4}
\usepackage{amsfonts}
\usepackage{amssymb}
\usepackage{graphicx}
\usepackage{subfigure}
\usepackage{color}
\usepackage{ulem}
\usepackage{bbold}
\usepackage{placeins}
\usepackage{soul,xcolor}
\usepackage{epstopdf}
\usepackage{amsmath}
\usepackage{cases}
\usepackage{braket}

\usepackage{lineno,hyperref}
\modulolinenumbers[5]

\usepackage{epsfig}
\usepackage{amsmath}
\usepackage{amsfonts}
\usepackage{graphicx}
\usepackage{color}
\usepackage{float}

\newcommand{\be}{\begin{equation}}
\newcommand{\ee}{\end{equation}}
\newcommand{\ba}{\begin{array}}
	\newcommand{\ea}{\end{array}}
\newcommand{\bea}{\begin{eqnarray}}
\newcommand{\eea}{\end{eqnarray}}

\def\bra#1{\left\langle #1\right|}
\def\ket#1{\left| #1\right\rangle}

\begin{document}
		\title{A quantum information theoretic quantity sensitive to the neutrino mass-hierarchy}
		
		\author{Javid Naikoo} \affiliation{Indian Institute of Technology Jodhpur, Jodhpur 342011, India}
		\author{Ashutosh Kumar Alok} \affiliation{Indian Institute of Technology Jodhpur, Jodhpur 342011, India}
		\author{Subhashish Banerjee} \affiliation{Indian Institute of Technology Jodhpur, Jodhpur 342011, India}
		\author{S. Uma Sankar} \affiliation{Indian Institute of Technology Bombay, Mumbai 400076, India}
		\author{Giacomo Guarnieri} \affiliation{Department of Optics, Palacky University, 17. listopadu 1192/12, 771 46 Olomouc, Czech Republic}
		\author{Christiane Schultze} \affiliation{University of Vienna, Faculty of Physics, Boltzmanngasse 5, 1090 Vienna, Austria}
		\author{Beatrix C. Hiesmayr} \affiliation{University of Vienna, Faculty of Physics, Boltzmanngasse 5, 1090 Vienna, Austria}
		
		
		\begin{abstract}
			    In this work, we derive a quantum information theoretic quantity similar to the
				Leggett-Garg inequality, which can be defined in terms
				of neutrino transition probabilities. For the case of
				$\nu_\mu \to \nu_e/\bar{\nu}_\mu\to\bar{\nu}_e$ transitions,
				this quantity is sensitive to CP violating effects as well as
				the  neutrino mass-hierarchy, namely which neutrino mass
				eigenstate is heavier than the other ones. The violation of
				the inequality for this quantity shows an
				interesting dependence on mass-hierarchy. For normal (inverted)
				mass-hierarchy, it is significant for $\nu_\mu \to \nu_e$
				($\bar{\nu}_\mu \to \bar{\nu}_e$) transitions. This is applied to the  two ongoing accelerator experiments T$2$K and NO$\nu$A as well as the future experiment DUNE.
		\end{abstract}


\maketitle

\section{Introduction}

One of the biggest open question is how a classical world could emerge out of the quantum world. In this connection the question arises are \textit{macroscopic superpositions}, e.g. the superposition of a Schr\"odinger cat in two distinct states, dead and alive, possible? From the technical point of view macroscopic superpositions are suppressed by decoherence, namely the (unwanted) coupling to the degrees of freedom of the environment, see e.g., Ref.~\cite{Xu}. Other approaches are so called spontaneous collapse models~\cite{Ghirardi} which propose an ontologically objective mechanism for a collapse in order to avoid the unobserved macroscopic superpositions. Recently, systems in high energy physics, including neutrinos, were studied and it was shown that these systems are  very suited to come up with a conclusive test~\cite{Catalina,Sandro,Kyrylo1,Kyrylo2,jan1}.

In this paper, we will tackle this question in a twofold way. We will define desired assumptions about the properties of a physical system and see whether those are valid for the three-flavor scenario of neutrino oscillation. For that purpose, we will derive a measurable quantum information theoretic quantity,  whose predictions satisfy an inequality.  Consequently, a violation of this inequality shows the conflict with the assumptions. We will further show that this quantity turns out to be also sensitive to two pertinent  questions within neutrino physics, namely the mass-hierarchy problem (whether $\nu_3$ mass-eigenstate is heavier or lighter than $\nu_1$ and $\nu_2$ mass-eigenstates) and how a possible violation of charge-conjugation--parity ($CP$) could be revealed. Thus our findings pave a road to tackle these questions from a quantum information theoretic perspective.

The mass-hierarchy problem refers to the ordering of the neutrino mass eigenstates ($\nu_1,\nu_2,\nu_3$) which is either $m_1<m_2<m_3$ (normal hierarchy or NH) or $m_3<m_1<m_2$ (inverted hierarchy or IH),
where $m_i$ is the mass of the state $\nu_i$. Solar neutrino data requires $m_2> m_1$. The violation of the combined discrete symmetries, charge-conjugation $C$ and parity $P$, is a well-studied topic since it relates to the cosmological question why we live in a universe dominated by matter and not anti-matter.

Leggett and Garg~\cite{Leggett} found an inequality based on the assumptions of \textit{macrorealism} and \textit{non-invasive measurability} which is considered as a test of macroscopic coherence~\cite{Emary,Kofler,Fritz}. Here, macrorealism means that the measurement of an observable $\hat{Q}$ of a macroscopic system reveals a well defined pre-existing value and non-invasive measurability states that, in principle, one determines this value without disturbing the future dynamics of the system. These assumptions lead to bounds on certain combinations of two time correlation functions $C(t_i, t_j) = \langle \hat{Q}(t_i) \hat{Q}(t_j) \rangle$, which may not be respected by quantum systems.  Since the notion of ``realism'' is often linked to the existence of  hidden variable theories, the violation of LGI nullifies the existence of such theories \cite{Suzuki, Huelga, Hardy}.

The well-known Bell inequalities could be considered as the spatial counterpart of Leggett-Garg inequality. Both inequalities capture, if violated, a contradiction with the assumptions on which these inequalities are based. Refs.~\cite{2flav} and~\cite{3flav} show the violation of Bell--type inequality in the context of two and three flavor neutrino oscillations~\cite{blasone}, respectively. In Ref.~\cite{Weiss} a method was proposed for probing the Leggett-Garg inequality within the framework of two flavor neutrino oscillations and it was shown that this inequality is violated in the MINOS experimental setup. Similar observations were reported recently for the Daya-Bay experiment in Ref.~\cite{Qiang}. Both these analyses were done assuming only two flavor neutrinos, which means they cannot be sensitive to  $CP$ violation. Further, these analyses can be applied only to those neutrino oscillation experiments which measure survival probabilities. In Ref.~\cite{Gangopadhyay} the authors studied the difference in the maximal violation of the Leggett-Garg inequality for a particular setup for two and three neutrino-flavor oscillations.

$CP$ violation in the quark sector is shown to be a crucial ingredient in the violation of a Bell inequality for entangled K-meson pairs~\cite{HiesmayrBI}, namely only due to the $CP$ violation the inherent nonlocality of the system is revealed for a certain setup.  An interplay between various facets of quantum correlations and $CP$ violation in the quark sector has been a topic of numerous interests, see for example \cite{Ancochea:1998nx,Hiesmayr:2011na,Banerjee:2014vga}. A theoretical analysis of the impact of entanglement on neutrino oscillation wavelength and its possible implications on the mass squared splittings can be found in Ref.~\cite{BKayser1}. How the nature of neutrinos, e.g. if they are Majorana or Dirac particles, can be revealed by a geometric phase was discussed in Ref.~\cite{HiesmayrMajoranaDirac}.

In this work, we develop an inequality similar to the Leggett-Garg inequality which will invoke macrorealism and stationarity. The advantage of this bounded quantity over the Leggett-Garg inequality is its experimental feasibility since all terms of the quantum information theoretic expression can be related to measurable probabilities. This allows a direct relation to the  ongoing experiments namely NO$\nu$A~\cite{Patterson,Adamson} and  T2K~\cite{Abe} and the upcoming experiment DUNE~\cite{Acciarri}.


The paper is organized as follows. We  discuss briefly the Leggett-Garg inequalities and derive a quantity that we can apply to the neutrino system. We include both matter effects (which turn out to be crucial) as well as $CP$ violating effects. We then compute the maximum of the Leggett-Garg parameter for the current accelerator neutrino oscillation experiments T2K  and NO$\nu$A  as well as future experiment DUNE followed by the conclusion.

\section{A quantum information theoretic quantity}

Consider a generic finite-dimensional quantum system which is assumed to have a finite transition probability $P_{i\to j}(t)$ to change from state $\ket{i}$ to state $\ket{j}$ during its evolution up to time $t$.
At that time $t$, imagine that one tries to ascertain whether the system is in a certain target state. This corresponds to performing a measurement of an observable $\hat{Q}$ which is dichotomic, i.e. whose possible outcomes are $\pm 1$ depending on whether the system is or is not found in the target state.
The degree of coherence of the dynamics is then captured by the autocorrelation function
\begin{eqnarray}
C(t_i,t_j) &:=&\langle \hat{Q}(t_i) \hat{Q}(t_j) \rangle\;  =\; Tr[ \{\hat{Q}(t_i), \hat{Q}(t_j)\}\rho(t_0)]\;,
\end{eqnarray}
where $\{\}$ denotes the anticommutation operation and the average is taken w.r.t. the initial state $\rho(t_0)$ and we assumed the time ordering $t_0\leq t_i\leq t_j$.
According to any \textit{realistic} description, any measurement would have revealed nothing more than a pre-existing state of the system, and thus the dynamical random variable corresponding to its possible outcome represents a definite value at any instant of time \cite{Leggett}. This would imply a bound on the following linear combination of autocorrelation functions
\begin{equation}\label{eq:Bound1}
K_3 =\; C(t_1, t_2) + C(t_2, t_3) - C(t_1, t_3)\; \le\; 1,
\end{equation}
where we have assumed by construction $t_1 \leq t_2 \leq t_3$.
It is worth stressing that Eq.~\eqref{eq:Bound1} can be derived from realism alone, without any assumption of \textit{non-invasive measurement}~\cite{Waldherr}.
As stressed already in the Introduction, the aim of the present work is to investigate the coherence of three-flavor neutrino dynamics. The assumption of non-invasive measurements, however, does not have a clear macroscopic counterpart, and, moreover, any insofar measurement of neutrinos requires destroying them, thus rendering this assumption ill-posed. For these reasons, we decide to go along the line of \cite{Huelga} and characterize the eventual coherence of neutrino's dynamics by testing the predictions of a fully quantum-mechanical description against a sub-class of realistic theories, namely the \textit{stationary} ones.

According to the latter condition, the autocorrelation function $C(t_i,t_j)$ actually only depends on the time-difference $t_j-t_i$, this leading to the following simplification of Eq. \eqref{eq:Bound1} \cite{Huelga,Zhou2}, denoted as Leggett-Garg-type inequalities:
\begin{equation}\label{eq:LGtI}
K_3|_{stat}\; =\; 2 C(0, t) - C(0,2t) \;\le\; 1,
\end{equation}
where we have assumed further that $t_1 = 0$ and that $t_2 - t_1 = t_3 - t_2 \equiv t$.

It is finally worth stressing that, according to Ref.~\cite{Emary}, a proper derivation of Eq.\eqref{eq:LGtI} can be obtained provided that: (i) the system is prepared in a given target state $\ket{j}$ at the initial time $t=0$; (ii) the system undergoes a \textit{unitary} or \textit{Markovian} evolution; (iii) the conditional probabilities $P(j,t+\tau | j,\tau)$ are time-translation invariant, i.e., $P(j,t+\tau | j,\tau) = \mathcal{P}(j,t|j,0)$.
The validity of (i) will be granted by the assumption to start from a well-defined flavor eigenstate of neutrinos and (ii) descends automatically from the fact that, as discussed below, the dynamics is unitary.  It is not difficult to  show that the time-translation invariance of conditional probabilities holds here.

\section{Neutrino evolution in vacuum and in a constant matter density}
\label{sec:QM}


Here, we discuss  the basic  elements of the quantum mechanical description of  neutrino dynamics. We will next employ them to calculate the autocorrelation function $C(t_i,t_j)$ of a particular dichotomic observable and compare the result with that of a stationary and realistic theory through the test of violations of the Leggett-Garg-Type Inequality (LGtI) given in Eq.~\eqref{eq:LGtI}.

A general neutrino state in flavor basis ($\ket{\nu_{\alpha}}$, $\alpha=e,\,\mu,\,\tau$) is given by the superposition
$\ket{\Psi (t)} = \nu_{e}(t) \ket{\nu_{e}} + \nu_{\mu}(t) \ket{\nu_{\mu}} + \nu_{\tau}(t) \ket{\nu_{\tau}}$.
The same state can be expressed in terms of mass eigen-basis ($\ket{\nu_{i}}$, $i=1,\,2,\,3$),
$\ket{\Psi (t)} = \nu_{1}(t) \ket{\nu_{1}} + \nu_{2}(t) \ket{\nu_{2}} + \nu_{3}(t) \ket{\nu_{3}}$.
The coefficients in the two representations are connected by a \textit{unitary} matrix
\be
\nu_\alpha(t) = \sum_{i=1,2,3} U_{\alpha i}\, \nu_{i}(t),
\ee
where $U_{\alpha i}$ are the elements of a 3 $\times$ 3 unitary PMNS (Pontecorvo-Maki-Nakagawa-Sakata) mixing matrix $U$ parametrized by three mixing angles
($\theta_{12},\,\theta_{23},\,\theta_{32}$) and a $CP$ violating phase $\delta$.  Experimental values for the PMNS mixing matrix are incorporated from the particle data group \cite{pdg}.

The coefficients in flavor basis at different times, i.e., $\nu_\alpha(t)$ in terms of $\nu_\alpha (0)$, are given by
\begin{equation}
{\nu}_\alpha(t) = {U}_{f} ~\nu_\alpha (0).
\end{equation}
In the absence of matter effects, the elements of the \textit{flavor} evolution matrix ${U}_{f}$ are functions of the PMNS matrix parameters.

When neutrinos propagate through matter with constant (electron) density $N_e$, because of a feeble interaction with electrons, the Hamiltonian, which is diagonal in the mass-eigen basis, $H_m = diag[E_1, E_2, E_3]$, picks up an interaction term, diagonal in flavor basis, $V_f = diag[A, 0, 0]$. Here $A=\pm \sqrt{2} G_F N_e$, the matter potential and $G_F$ is the Fermi coupling constant.  The flavor evolution operator for constant matter density takes the form \cite{Tommy,TTommy}
\begin{equation}
U_{f}(L) = \phi ‎‎\sum_{n=1}^{3}  \frac{e^{-i \lambda_n L}}{3\lambda_{n}^{2} + c_1} \left[ (\lambda_{n}^{2}  + c_{1}) \mathbf{I} + \lambda_n \tilde{T} + \tilde{T}^2 \right],
\label{ufmatter}
\end{equation}
where $\phi = e^{-i \frac{Tr\mathcal{H}_m}{3} L}$, $c_1 = det(T) Tr(T^{-1})$, $T=\mathcal H_m-(Tr\mathcal H_m)\mathbf{I}/3$ and the Hamiltonian in mass basis is
\begin{equation}
\mathcal{H}_m = H_m + U^{-1} V_f U.
\end{equation}
The matrix $\tilde{T} = U T U^{-1}$, where  $U$ is the PMNS mixing matrix. $T$ is a Hermitian matrix~\cite{Tommy} with eigenvalues $\lambda_n ~(n=1, 2, 3)$.  The effect of  variable matter density on neutrino oscillations can be found in Refs.~\cite{Tommy2001, Tommy2002}.

\section{Leggett-Garg type inequality for neutrinos}

Here we derive an analytic expression based on Eq.~\eqref{eq:LGtI} for the case of three-flavored neutrinos, when the latter are prepared in a specific flavor state $\ket{\nu_\alpha}$ ($\alpha = e/ \mu/\tau$). The problem can be dealt with by choosing appropriate dichotomic observable $ \hat{Q(t)} = 2 \ket{\nu_\beta(t)} \bra{\nu_\beta(t)} - \mathbb{1}$~\cite{Budroni}, with the completeness condition $\sum_{\beta=e, \mu, \tau} | \nu_\beta \rangle \langle \nu_\beta |=\mathbb{1}$. The latter therefore  makes quantitative the inquiry as to whether the neutrino is in flavor $\beta$ ($Q=+1$) or not ($Q=-1$) at a certain time instant $t$. It is worth mentioning here that the extent of violation depends on the number of levels $N$ and the number of projectors $M$. With just two projectors, the maximum quantum value (L\"uders bound) of the LG parameter $K_3$ is $3/2$, \textit{irrespective} of the dimensions of the system~\cite{Budroni, Budroni2, Lambert1}. In our case, $N=3$ and $M=2$.

Under the \textit{stationary} assumption, the autocorrelation functions $C(t_i, t_j) \equiv C(0,t)$ are straightforwardly found and can be compactly expressed in terms of the probability $P_{\alpha\to\beta}(t) = | \langle \nu_{\alpha}(t)|\nu_{\beta}(0) \rangle|^2$ as
\begin{equation}\label{C0t}
C(0,t) = 1 - 2 P_{\alpha\to\beta}(t).
\end{equation}
Note that both the survival probability $P_{\alpha\to\alpha}(t)$ and the transition probability $P_{\alpha\to\beta}(t)$ actually depend on many physical parameters such as the neutrino energy $E$, the mass square differences $\Delta_{ij} =  m_j^2 - m_i^2$, the matter potential $A$, the mixing angles $\theta_{ij}$ and the $CP$ violating phase $\delta$.
For clarity of notation, however, we will keep the dependence on all these parameters implicit except for the energy $E$, for reasons that will be clear in a short-while. Care, however, should be taken when moving from neutrinos to anti-neutrinos since both the $CP$ violating phase $\delta$  and the matter potential
$A$ reverse their signs~\cite{Nunokawa}.
The quantity of interest thus becomes
\begin{equation}\label{eq:MainResult}
K_3 =   1 + 2 P_{\alpha\to\beta}(2t,E) -  4 P_{\alpha\to\beta}(t,E),
\end{equation}
which shows its experimental feasibility, being clearly expressed only in terms of measurable quantities, i.e., survival and transition probabilities.

In light of its physical grasp, the following considerations on Eq.~\eqref{eq:MainResult} naturally follow.
Neutrino oscillation experiments typically operate in the ultra-relativistic regime and therefore the time-dependence in the probabilities $P_{\alpha\to\beta}(t,E)$ can be equivalently replaced by the length $L$ travelled by neutrinos \cite{Kim}.
An experimental verification of Eq.~\eqref{eq:MainResult} would thus require two detectors placed at $L$ and $2 L$, respectively.
The current experimental facilities, however, do not allow for such a setup.
It is nevertheless possible to bypass such obstacle by looking for matching energies $\tilde{E}$ satisfying the implicit equation
\begin{equation}\label{eq:tildeE}
{P}_{ \alpha \rightarrow \beta}(2L, E)\; \equiv\; {P}_{  \alpha \rightarrow \beta}(L, \tilde{E}).
\end{equation}
Let us remark here that of course this identification can lead to multiple solutions and both probabilities may not have the support on the same interval of energies.

\begin{figure*}[t]
	\centering
	\includegraphics[width=0.6\textwidth]{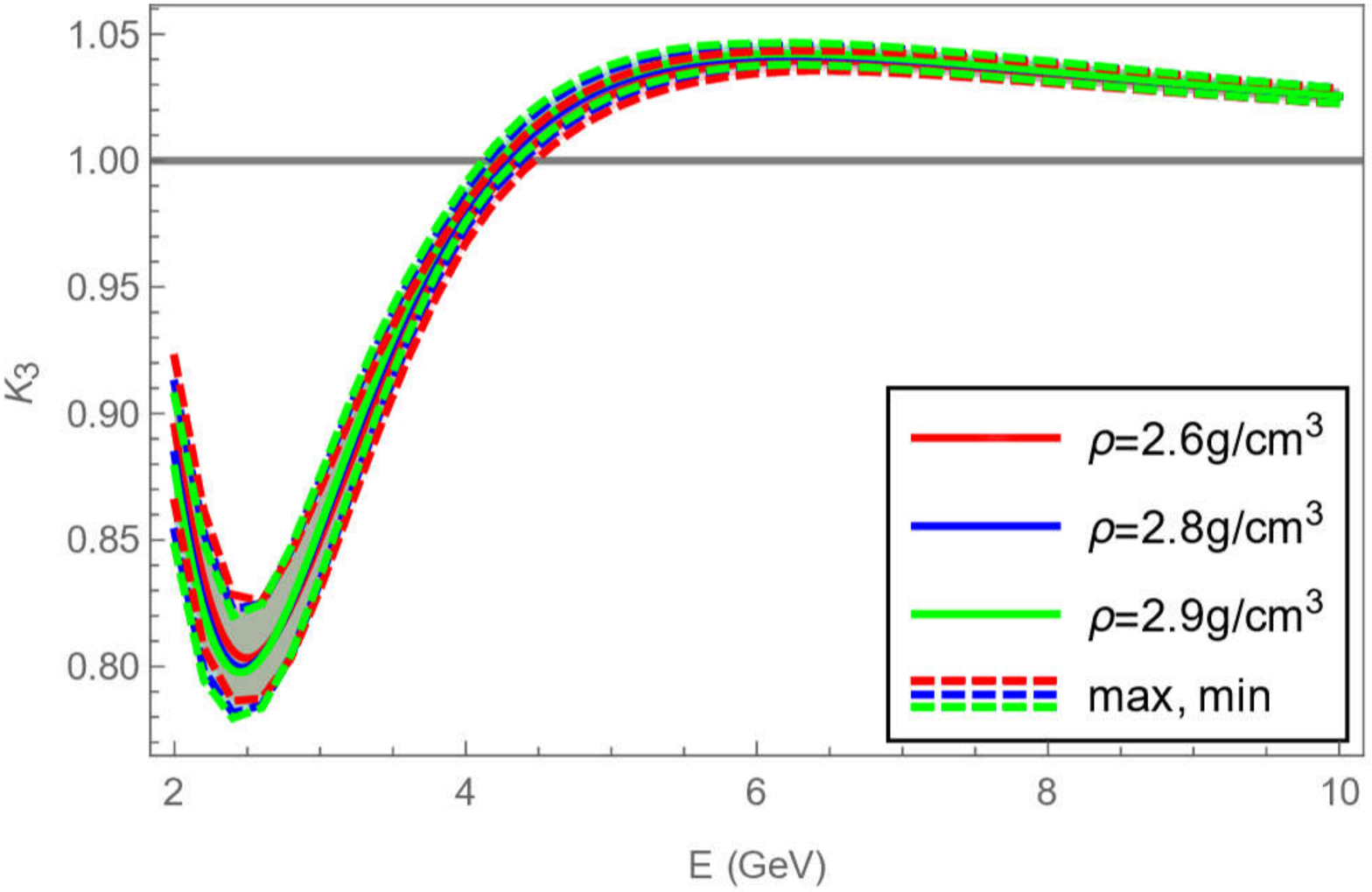}
	\caption{(Color Online) These graphs show the value of the quantity $K_3$, for different values of the matter density $\rho=2.6/2.8/2.9g/cm^3$ with the matter potential $A=A(\rho)$ in dependence of the energy $E$ and including the known errors for the other parameters, i.e. the three mixing angles $\theta_{ij}$ and the three mass differences $\Delta_{ij}$ given in Ref.~\cite{pdg}. Mean values of the three mixing angles $\theta_{ij}$ and the three mass differences $\Delta_{ij}$ are taken according to Ref.~\cite{pdg} as well. The (dotted) curves correspond to the mean values for which the lower and upper possible value for the errors were computed numerically. A length $L$ of 1300 km and a $CP$ violating phase $\delta$ of $0$ are considered.}
	\label{fig1}
\end{figure*}

\section{Applications and discussion}

Now we are ready to exploit our result Eq.~(\ref{eq:MainResult}) for neutrinos and antineutrinos and to connect it to the experiments via Eq.~(\ref{eq:tildeE}).  The experiments under consideration employ $\nu_{\mu}/\bar{\nu}_{\mu}$ beams as sources and study the survival probabilities $P(\nu_\mu\to \nu_\mu)(t,E)$ and $P(\bar{\nu}_\mu\to \bar{\nu}_\mu)(t,E)$ and also the transition probabilities $P(\nu_\mu\to \nu_e)(t,E)$ and $P(\bar{\nu}_\mu \to \bar{\nu}_e)(t,E)$. In this work,  we  concentrate on the transition probabilities, which are sensitive to matter effects and $CP$ violation in long baseline accelerator neutrino experiments. We evaluate
\begin{equation}\label{mutoe}
K_3 = 1 + 2 P_{ \mu \rightarrow e}(L,\tilde{E}) - 4 {P}_{ \mu \rightarrow e}(L,E),
\end{equation}
for both neutrinos and anti-neutrinos. In each case, the calculation is done
for both positive and negative values of $\pm\Delta_{31}$ and for a given matter potential $A$.

Let us first investigate the sensitivity of $K_3$ on the matter potential $A$ as a function of the energy $E$  by taking into account the errors in the mixing and  mass squared differences, which is displayed in Fig.~\ref{fig1}. For high energies the quantity $K_3$ depends less on the uncertainties in the experimental values and a clear violation is observed. The dependence of  the function $K_3$ on the experimental uncertainties is higher for lower energies and no violation is found below a certain energy.


\subsection{DUNE}

\begin{figure}[h]
	\centering
	\includegraphics[width=70mm]{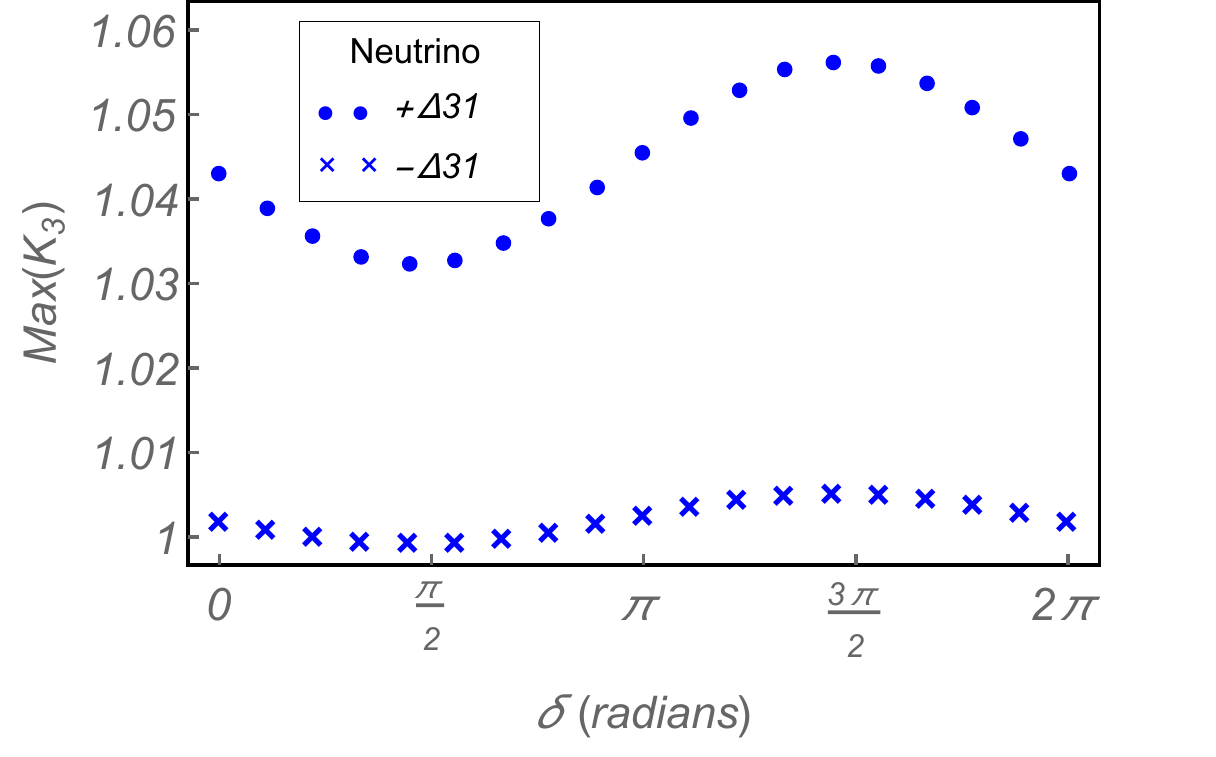}
	\includegraphics[width=70mm]{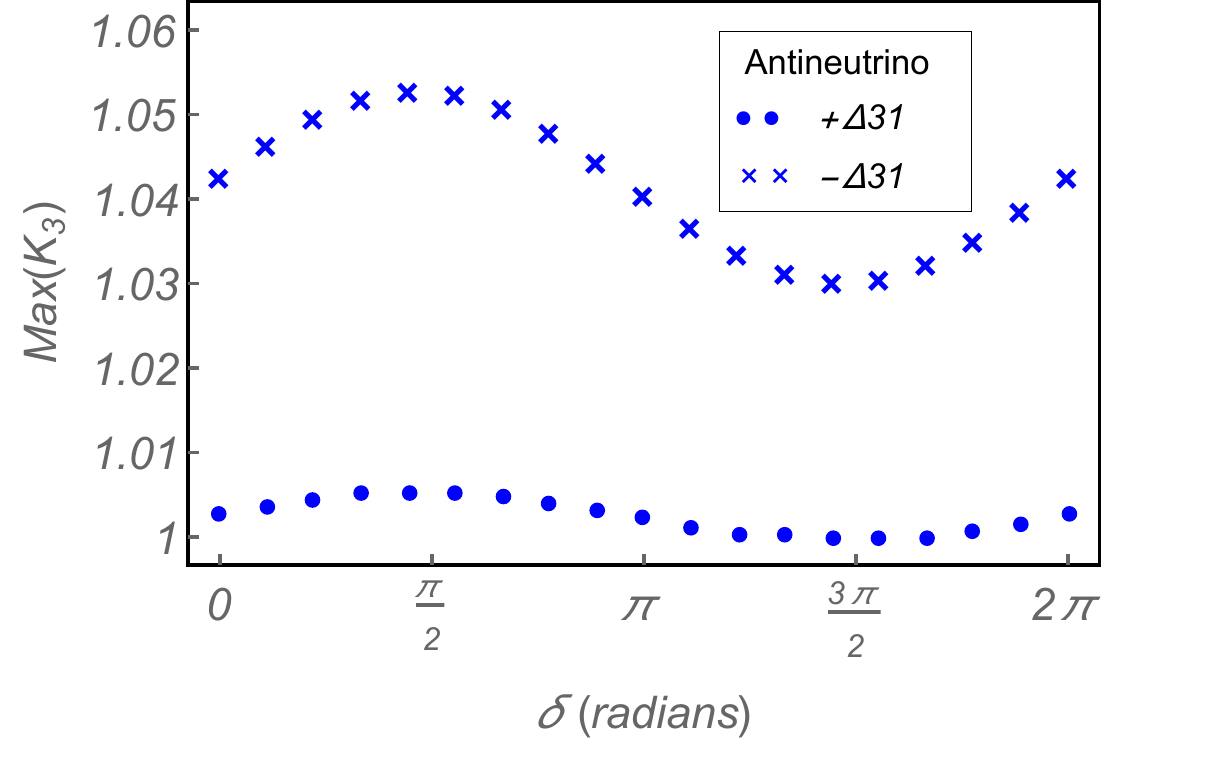}
	\caption{DUNE: Maximum of $K_3$ plotted against $CP$ violating phase $\delta$ in the presence of matter effects ($\rho=2.8g/cm^3$). Dotted and crossed curves correspond to the positive and negative signs of $\Delta_{31}$, respectively. Length $L$ is equal to 1300 km and the neutrino energy $E$ is varied between 2 GeV to 10 GeV.
		The corresponding range of $\tilde{E}$ is 1 to 5 GeV. The top and the bottom panels refer to neutrinos
		(positive mass density $+A$ and positive $CP$ violating phase $\delta$) and anti-neutrinos ($-A$, $-\delta$), respectively.}
	\label{dune}
\end{figure}

The DUNE facility located at Stanford Underground Research Laboratory at South Dakota will have access to a rather wide band energy $1-10$ GeV, thus allowing a test $K_3$ via the identification $E$ and $\tilde{E}$, Eq.~\ref{eq:tildeE}.
%
In Fig.~(\ref{dune}), we plot the maximum of $K_3$ over a certain energy interval $\Delta E$ which is identified with a certain energy interval $\Delta \tilde{E}$,
Eq.~\ref{eq:tildeE}, as function of the $CP$ violating parameter $\delta$. More explicitly, the maximization is performed for a given $A,\,\Delta_{31}$ at fixed values of the $CP$ violating phase $\lbrace\delta_k\rbrace$ and within the energy window of the experimental setup. It is also worth stressing that the value of $\tilde{E}$ is found by solving Eq. \eqref{eq:tildeE} after imposing the above constraints on the mass square difference $\Delta_{31}$ and $\delta = \delta_k$.

\begin{figure}[h]
	\centering
	\includegraphics[width=70mm]{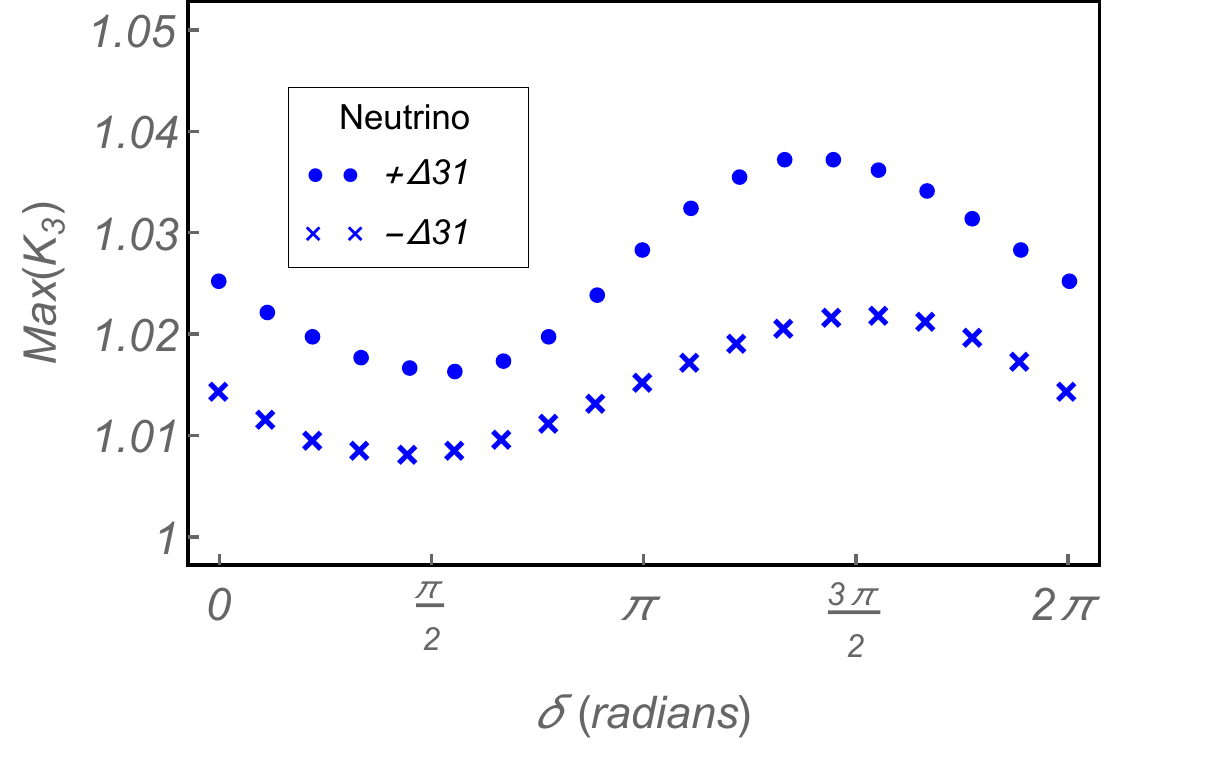}
	\includegraphics[width=70mm]{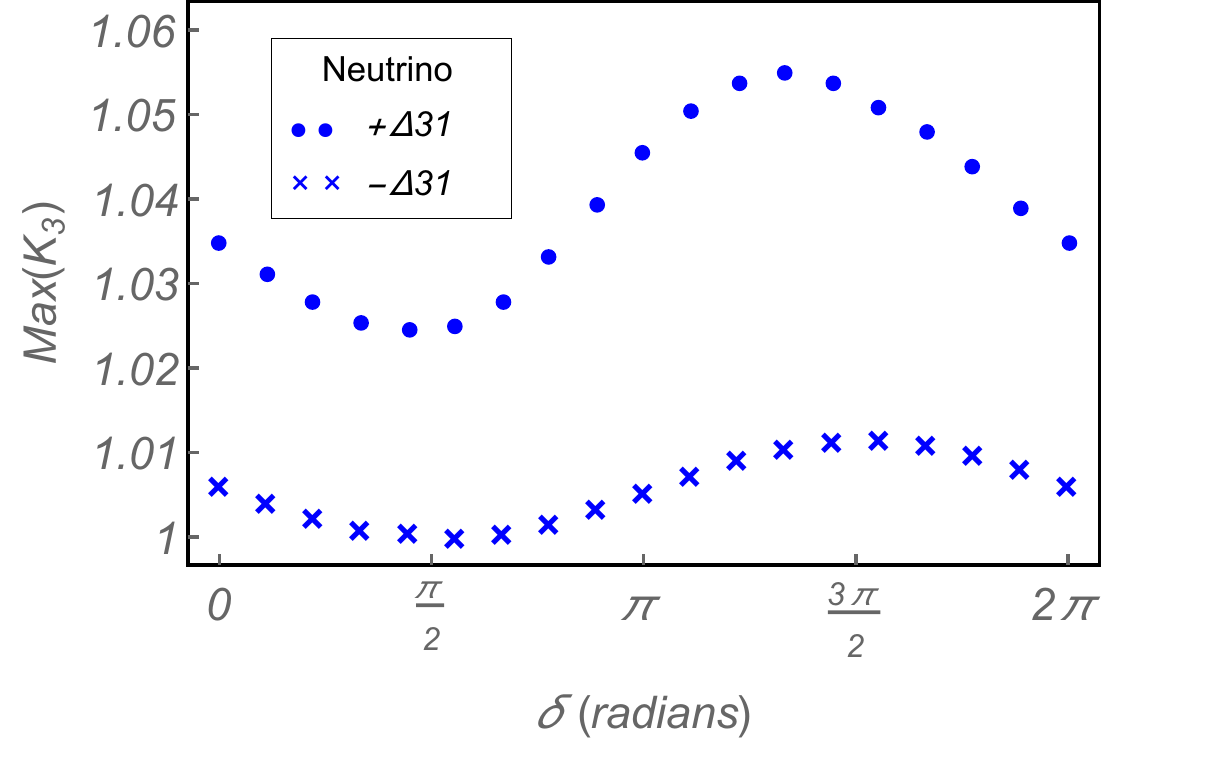}
	\caption{T2K and NO$\nu$A: Maximum of the parameter $K_3$ plotted vs $CP$ violating phase $\delta$
		for T2K (top panel) and NO$\nu$A (bottom panel). Dotted and crossed curves correspond to the positive and negative signs of $\Delta_{31}$, respectively. Length L is 295 km and 810 km for T2K and for NOvA, respectively. The energy $E$ is varied between 1 GeV to 2 GeV for T2K with $\tilde{E}$ between 0.1 GeV to 1 GeV, while for NO$\nu$A $E$ is taken between 1.5 GeV to 5 GeV with $\tilde{E}$ between 0.1 GeV to 3 GeV.}
	\label{t2knova}  
\end{figure}

Two interesting features stand out from Fig.~\ref{dune}. Maximum value of $K_3$
for neutrinos is larger (smaller)  for $\delta$ in the lower (upper) half plane. This is a
reflection of the dependence of $P(\nu_\mu \to \nu_e)$ on $\delta$. More interestingly, the $K_3$ violation in neutrinos is nearly an order of magnitude larger for the case
of positive $\Delta_{31}$ compared to the case of negative $\Delta_{31}$. The situation is
reversed for anti-neutrinos. These plots indicate  that one should attempt to measure
$K_3$ using neutrino data for which $\Delta_{31}$ is positive, whereas the anti-neutrino
data should be used for such an attempt when $\Delta_{31}$ turns out to be negative.

\subsection{T2K and NO$\nu$A}

The current accelerator neutrino experiments, T2K and NO$\nu$A, both have rather
narrow energy bands. For T2K, the baseline is 295 km and the energy range is
$0.5 - 2$ GeV. For NO$\nu$A, the corresponding numbers are 810 km and $1-4$ GeV.
These experiments were planned before the mixing angle $\theta_{13}$ was measured to
be moderately large. Their neutrino beams were designed to be narrow band beams to
suppress the backgrounds.  Fig.~\ref{t2knova} depicts the maximum of $K_3$ for T2K and NO$\nu$A experiments for the parameters (energy, baseline and matter density) of the latest neutrino runs.  The corresponding plots for anti-neutrino run can be obtained from  Fig.~\ref{t2knova} by the maps $\rm NH \leftrightarrow IH$ and $\delta \leftrightarrow 2\pi - \delta$, as in the case of DUNE.

We see that the plots for positive and negative values of $\Delta_{31}$ are well separated
in the case of NO$\nu$A whereas the separation is much less for the case of T2K. This is a consequence
of the matter effect for T2K being much less than that of NO$\nu$A.
It is clear from the above discussions, that owing to the wide energy band, DUNE experiment is suitable for studying $K_3$ and its sensitivity to the neutrino mass-hierarchy and $CP$ symmetry violations.  Further, the violation of the LGtI is significant in $\nu_\mu \to \nu_e$ ($\bar{\nu}_\mu \to \bar{\nu}_e$) transitions for normal (inverted) mass-hierarchy.

\section{Conclusion}

In this work we have evaluated $K_3$, a quantity related to assumptions about the  macroscopic realism of a physical system, in the context of three-flavor neutrino oscillations. We found  a closed expression for the $K_3$ function in terms of neutrino oscillation probabilities and, consequently, a direct relation to the experiments.

From the quantum information theoretic perspective we have found that for initial muon-neutrino states and a broad energy band available the quantity $K_3$ is violated in nearly all cases and independent of the $CP$ violating parameter. However, there are strong differences if electron-neutrinos or electron-anti-neutrinos are considered in the final state and if we have the scenario of normal or inverted hierarchy, Figs.~\ref{dune} and \ref{t2knova}. The violation is more prominent for $\nu_\mu \to \nu_e$ ($\bar{\nu}_\mu \to \bar{\nu}_e$) transitions for normal (inverted) mass-hierarchy. The huge difference in the behaviour of the function $K_3$ is due to the matter effect. This in turn depends strongly on the uncertainties of the parameters in the three-flavor neutrinos evolution and the energy, see Fig.\ref{fig1}.

From the experimental point of view we have found a combination of experimentally measurable quantities, probabilities, that reveal particular information about the underlying physics such as about the ordering of masses. As can be seen from the Figs.~\ref{dune} and \ref{t2knova} the experiment DUNE is appropriate to take practical advantage of this.

The present work therefore paves the way for interesting cross fertilization of ideas from particle physics to address foundational concepts of quantum mechanics and an experimental road to address deeper questions in  the neutrino system.

\section*{Acknowledgments}

The work of all authors except S.U.S. was supported by DST India-BMWfW Austria Project, based Personnel Exchange Programme for 2017-2018. S.B., J.N., and A.K.A. acknowledge Khushboo Dixit for help in the numerics.   G.G. acknowledges the support of the Czech Science Foundation (GACR)
(grant No. GB14-36681G). B.C.H. and Ch.S. acknowledge gratefully the Austrian Science Fund (FWF-P26783).

\end{document}